# Predicting Triple Scoring
# with Crowdsourcing-specific Features

## The fiddlehead Triple Scorer at WSDM Cup 2017

**Masahiro Sato**
Fuji Xerox Co., Ltd.
sato.masahiro@fujixerox.co.jp

**ABSTRACT**

The Triple Scoring Task at the WSDM Cup 2017 involves the prediction of the relevance scores between persons and professions/nationalities. The ground truth of the relevance scores was obtained by counting the vote of seven crowdworkers. I confirmed that features related to task difficulty correlate with the discrepancy among crowdworkers' judgement. This means such features are useful for predicting whether a score is in the middle or not. Hence, the features were incorporated into the prediction model of the crowdsourced relevance scores. The introduced features improve the average score difference of the prediction. The final ranking of my prediction was 4th for average score difference and 12th for both accuracy and Kendall's tau.

## 1. INTRODUCTION

Knowledge bases are important for information search. They can be extracted from open data sources like Wikipedia and Freebase, which are created by collaborative works of crowds. Relevance or importance of the facts in knowledge bases differs among the facts. This is especially true for type-like relations like persons and professions. The ground truth data of such relevance can be obtained by crowdsourcing, which utilizes the power of crowds, as in the Triple Scoring Task at the WSDM Cup 2017 [1, 2].

This work is conducted as part of the Triple Scoring Task at the WSDM Cup 2017 [1, 2]. The task objective is to predict relevance scores of type-like relations between persons and professions/nationalities. The relevance score is the sum of seven crowdworkers' votes on whether they think that each profession/nationality is primary for the specific person. Hence, the score ranges from 0 to 7 by an integer. The crowdsourcing procedure is described in [4]. The contest organizers provided Wikipedia sentences, which include annotations of persons, although other data could be used. Small portions of relevance scores were also provided as training data. The prediction for test data was conducted on the evaluation platform TIRA [5], and the persons included in the test data were not revealed to the participants. The performance was evaluated by three measures: Average Score Difference (ASD), Accuracy (the percentage of predictions differing from the truth by at most 2), and Kendall's Tau (the ratio of discordant to concordant pairs on score orders). The details of the contest are described in [1, 2].

When predicting relevance scores obtained by crowdsourcing, the characteristics of crowdworkers' behaviors should be considered. The accuracy of crowdsourced results depends on task difficulty [3]. As the relevance score of the Triple Scoring Task is the sum of seven crowdworkers' votes, the difficult task might cause discrepancy among crowdworkers' judgements and the score might be in the middle.

In this competition, I have focused on extracting features related to task difficulty. Crowdworkers judge the relevance of multiple professions/nationalities for each person; hence, the difficulty of the task depends on the number of professions/nationalities. In addition, the popularity of the person or the familiarity of the professions/nationalities should facilitate the judgment. Hence, I have extracted these features from the data provided for the WSDM Cup. The correlation between these extracted features and the judgment discrepancy of crowdworkers were investigated. I incorporated these crowdsourcing-specific features into the prediction model of relevance scores, in addition to the relevance features similar to those introduced in the previous work [4].

The outline of this paper is as follows. The next section presents the details of my approach. Section 3 shows the evaluation results. Finally, I conclude this paper in Section 4.

## 2. APPROACH

I have combined the crowdsourcing-specific features (CS) and relevance features (Rel). Table 1 summarizes the features. The details of each feature are explained in Subsections 2.2 and 2.3. Regression trees were used for the final prediction. I used only the data provided by the contest organizers as shown in Table 2. The details of the data are described in [1, 2]. I used the R language in the implementation of my code.

**Table 1. Used Features**

| Type | Name | Definition |
|------|------|------------|
| CS | Person Popularity | # of wikisentences with the person's annotation |
| CS | Profession/Nationality Familiarity | # of persons with that profession/nationality |
| CS | Candidate Option | # of profession/nationality candidate for the person |
| Rel | Exact Count | # of exact phrases of profession/nationality in wikisentences |
| Rel | Weighted Keywords Count | # of related keywords weighted by 1/rank with tf-idf ranking |
| Rel | Logistic Regression Estimates | Estimates of binary classification for each profession/nationality |

**Table 2. Used Data**

| Name | Description |
|------|-------------|
| profession.kb | All professions for a set of 343,329 persons |
| profession.train | Relevance scores for 515 tuples (pertaining to 134 persons) from profession.kb |
| nationality.kb | All nationalities for a set of 301,590 persons |
| nationality.train | Relevance scores for 162 tuples (pertaining to 77 persons) from nationality.kb |
| professions | 200 different professions from professions.kb |
| nationalities | 100 different nationalities from nationalities.kb |
| persons | 385,426 different person names from the two .kb files and their Freebase ids |
| wiki-sentences | 33,159,353 sentences from Wikipedia with annotations from 385,426 persons |

## 2.1. Preprocessing

First, I linked sentences from Wikipedia to each person. "wiki-sentences" include annotations of persons (enclosed by brackets [] in the data). I used these annotations and removed them from the sentences after linking. For each person, linked sentences were collected and aggregated. Some sentences include multiple persons and were assigned to multiple persons. Terms in sentences were lower-cased. One-letter terms, digit-only terms, and English stopwords [6] were removed from the sentences.

Next, I extracted keywords for each profession/nationality. Some persons have only one profession in "profession.kb" or only one nationality in "nationality.kb". I regarded these pesrsons as positive examples for each profession/nationality (similar to [4]). I counted the number of occurrences of each term in the sentences of the persons in the positive examples. Then I computed the tf-idf of each term. The idf values were based on the whole sentences, that is, $1/\log(\# $ in the whole sentences). The terms were ranked according to their tf-idf values and the top 200 terms (keywords) for each profession/nationality were kept for preceding feature engineering.

## 2.2. Crowdsourcing Features

The difficulty of a crowdsourcing task affects the task result [3]. The difficulty could cause discrepancy in relevance judgment and the relevance scores of difficult judgements tend to be in the middle (namely, around three or four in 0-7 scale). I extracted three features related to the task difficulty. The first feature is the person popularity, which is defined as the number of wikisentences linked to that person. The more times the person is mentioned on Wikipedia, the more popular the person must be. Crowdworkers can easily answer the profession/nationality of a popular person, although the crowdworkers could resort to Wikipedia if they do not know the person. The second feature is the Profession/Nationality familiarity, which is the number of persons who have that profession/nationality in "profession.kb" or "nationality.kb". If there are many persons who have the same profession/nationality, then the profession/nationality is highly familiar. Crowdworkers have difficulty in judging unfamiliar profession/nationality. The third feature is the candidate option, which is the number of professions/nationalities for each person in "profession.kb" or "nationality.kb". Crowdworkers judged whether each profession/nationality was primary or not. If the person has many professions/nationalities, crowdworkers will have a hard time to complete the task.

I investigated the correlation between judgment discrepancies of relevance scores and the three features discussed previously. The judgment discrepancy was defined as the number of crowdworkers who judged differently from the majority (0 for relevance scores 0 or 7, 1 for relevance scores 1 or 6, and so forth). The relevance scores in "profession.train" were used for the analysis.

The number of candidate options was positively correlated with the discrepancy. Figure 1 shows the correlation between the judgment discrepancy averaged per person and the number of candidate professions. The darker dots in the figure denote the overlap of the data in the same coordinate points. The blue line shows the fit of the linear regression. Pearson's correlation coefficient of the data is 0.21. The positive correlation is statistically significant (p-value = 0.016). As the number of the candidate professions of the person increases, the discrepancy expands.

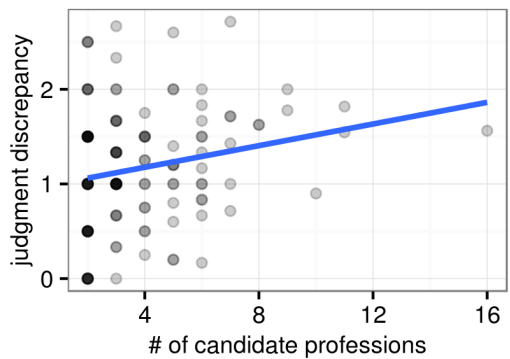

**Figure 1. Correlation between the judgment discrepancy and the number of candidate professions ("candidate option" feature).**

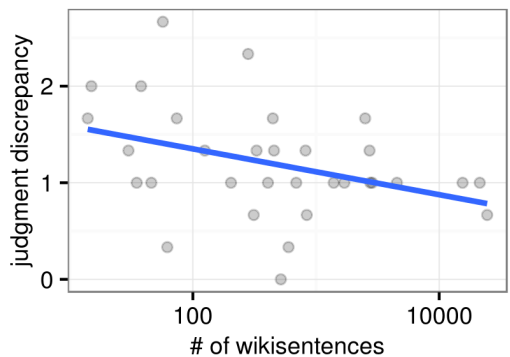

**Figure 2. Correlation between the judgment discrepancy and the number of wikisentences ("person popularity" feature) for persons with three profession candidates.**

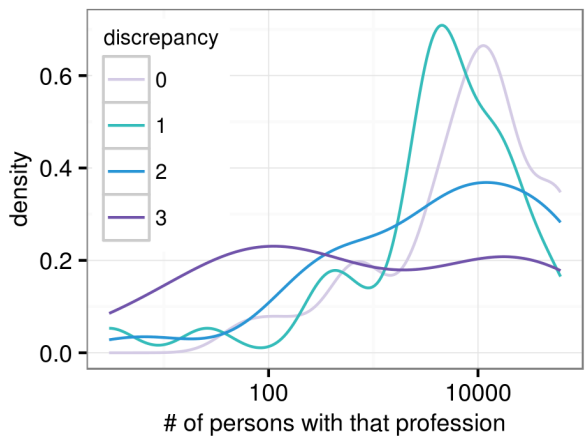

**Figure 3: Distribution of the number of persons according to profession ("profession familiarity" feature). The densities of the data with each discrepancy are plotted separately.**

The person popularity was negatively correlated with the discrepancy. Given that both discrepancy and popularity were positively correlated with the number of candidate professions, I compared the discrepancy for the same number of candidates. Figure 2 shows the result for the persons with three candidate professions. The x-axis is log-scaled. Pearson's correlation coefficient of the data is -0.37. The negative correlation is statistically significant (p-value = 0.019). The judgment discrepancy decreases for popular persons, presumably because crowdworkers can judge the profession of famous persons easily. I have analyzed the persons with several number of candidate professions, and found similar negative correlations.

The familiarity of profession/nationality was negatively correlated with the discrepancy. The familiarity distributions for each discrepancy are shown in Figure 3. The data with low discrepancy have higher familiarity compared with the data with high discrepancy. Familiar professions might be easier to judge for crowdworkers.

These features are useful for predicting relevance scores because higher judgement discrepancy means that scores are more in the middle.

### 2.3. Relevance Features

I used three features for profession/nationality relevance. The first feature is the exact count, which is the number of occurrences of profession/nationality name (e.g., American Football Player, Voice Actor) in wikisentences associated with each person. The match was case-insensitive. The second feature is a weighted keyword count. It is the sum of the 1/rank weights of keywords appearing in the associated sentences of each person for each profession/nationality (the same as "count profession words" in [5]). The rankings of keywords were based on tf-idf as mentioned in subsection 2.1. The top 20 keywords were used for this calculation.

The third feature is the logistic regression estimates (similar to binary classifier described in [4] with some differences). The top 50 keywords for each profession/nationality were used as features of the classifiers. Positive examples to train the classifier include the data for persons who have only one profession/nationality. Negative examples were randomly sampled from the data of the persons with another profession/nationality and the numbers of negative samples were set to ten times the numbers of positive samples. After learning the model for each profession/nationality, predicted scores on whether each profession/nationality was primary or not were calculated for persons with multiple professions/nationalities. These predicted scores (from 0 to 1) were then used as features "Logistic Regression Estimates" features to predict relevance scores. I used Liblinear [7] for this feature extraction. The R interface to Liblinear is provided with LiblineaR package [8]. The LiblineaR parameters were set to the default values, i.e., the model was L2-regularized logistic regression and the cost of constraints violation (inverse of a regularization constant) was 1.

## 2.4. Prediction Models

I constructed final prediction models using the features described in Subsections 2.2 and 2.3. For weighted keyword counts and logistic regression estimates, the values normalized with the maximum value for each person were also added as features (therefore, eight features in total). I used regression trees (rpart package [9]) for prediction. The relevance scores for training data were obtained from "profession.train" and "nationality.train". I used the default parameters of rpart. I tested several values for the complexity parameter, where default 0.01 was found to be the best.

The relevance scores for every person-profession pair in "profession.kb" and every person-nationality pair in "nationality.kb" were predicted by the models (excepting the persons with only one profession or one nationality). I assumed that the test data in TIRA would be sampled from these pairs. I added predicted relevance scores (rounded to an integer value) to profession.kb and nationality.kb and uploaded them to TIRA as reference data. Note that I have not truncated the score range within 2 and 5, and the prediction ranged from 0 to 7. The software I submitted was programmed to consult these files and return the predicted scores. In case the person-profession/nationality pairs were not included in the reference data, the software returned 4 as an answer.

## 3. EVALUATION RESULTS

I evaluated my models via 10-fold cross-validation using the relevance scores of "profession.train" and "nationality.train". The official results of submitted software were announced after the competition as shown in Table 3. Although my final submission was based on both the crowdsourcing features and relevance features, the cross validation results with only relevance features are also shown for the sake of reference. Adding CS for the prediction of the profession relevance improved ASD, while it slightly decreased accuracy and did not affect Kendall's tau. CS did not affect the prediction of nationality. The number of training tuples for nationality is less than one-third of that for profession, and incorporating the crowdsourcing-specific features might need more training data.

The results of the final test data were worse than the expected from the offline cross-validation. The possible reason is that my software could not properly encode some special characters especially of non-English names. The final ranking of my prediction was 4th for ASD and 12th for both accuracy and Kendall's tau.

**Table 3. Evaluation results. The results of training data cross-validation for profession/nationality are separated with "/". The results of test data are combinations of both profession and nationality relevance prediction.**

| Condition | ASD | Accuracy | Kendall tau |
|---|---|---|---|
| Train (only Rel) | 1.606/ 1.625 | 0.800/ 0.763 | 0.333/ 0.382 |
| Train (with CS) | 1.591/ 1.625 | 0.796/ 0.763 | 0.333/ 0.382 |
| Test (with CS) | 1.70 | 0.73 | 0.40 |

## 4. CONCLUSIONS

I introduced the crowdsourcing-specific features concerning the difficulty of the task for crowdworkers. The additions of these features to the conventional relevance features improved the average score difference of the prediction of crowdsourced relevance scores.

The correlation between the crowdsourcing-specific features and the relevance scores might suggest the need of careful interpretation of the crowdsourced relevance scores. The obtained scores could be a mixture of the real relevance and the side effect of task difficulty. To construct accurate knowledge bases, investigating the behaviour patterns of crowdworkers and extracting the real relevance would be an interesting future research.

## ACKNOWLEDGMENTS
I thank the developers of the valuable libraries used in this work. I appreciate the kind support of the contest organizers.